\documentclass{optica-article}
\journal{opticajournal} 
\articletype{Research Article}
\usepackage{graphicx} 
\usepackage{float}
\usepackage{geometry}
\usepackage{amsmath}
\usepackage{subcaption}
\usepackage{adjustbox}
\usepackage{comment}
\usepackage{amssymb}
\usepackage{booktabs}
\usepackage{siunitx}
\usepackage{hyperref}
\usepackage{float}
\usepackage[numbers]{natbib}
\usepackage{subcaption}
\usepackage{physics}
\usepackage{color}
\usepackage{cite}
\usepackage{siunitx}

\usepackage{tikz}
\usepackage{circuitikz}
\usetikzlibrary{positioning, shapes.geometric}
\usetikzlibrary{decorations.pathmorphing}
\usetikzlibrary{calc}
\usepackage{lineno}
\newlength{\figw}
\colorlet{bandfill}{red!70}
\colorlet{conncolor}{red!70}
\def\bandopacity{0.15}
\def\funnelopacity{0.10}
\def\bandfrac{0.13}       

\def\PanelLx{49.41}\def\PanelLxr{270.56}   
\def\PanelRx{308.45}\def\PanelRxr{530.32}  
\def\TopYb{256.14}\def\TopYt{408.68}        
\def\BotYb{45.57}\def\BotYt{197.87}         
\newcommand{\shadeband}[6]{%
  \fill[bandfill,opacity=\bandopacity]
    ({#1+(#2-#1)*#5},#3) rectangle ({#1+(#2-#1)*#6},#4);}
\newcommand{\zoomcol}[6]{%
  \pgfmathsetmacro{\br}{#5 + (#2-#1)*\bandfrac}
  \fill[bandfill,opacity=\funnelopacity]
    (#1,#3) -- (\br,#3) -- (#2,#4) -- (#6,#4) -- cycle;}

\begin{document}

\title{Precise Photon Arrival Time Measurement via Time to Frequency Demultiplexing}

\author{Usman Khan,\authormark{1,2} Santosh Kumar,\authormark{1,2} Daniel López,\authormark{1,2} Siqun Deng,\authormark{1,2} Malvika Garikapati,\authormark{3} and Yuping Huang\authormark{1,2,*}}

\address{\authormark{1}Department of Physics, Stevens Institute of Technology, 1 Castle Point Terrace, Hoboken, New Jersey, 07030, USA\\
\authormark{2}Center for Quantum Science and Engineering, Stevens Institute of Technology, 1 Castle Point Terrace, Hoboken, New Jersey, 07030, USA\\
\authormark{3}Quantum Computing Inc, 5 Marine view plaza, Suite 214,
Hoboken, New Jersey, 07030, USA}

\email{\authormark{*}yhuang5@stevens.edu} 

\begin{abstract*}
We demonstrate a nonlinear-optics approach to precise measurement of photon arrival time, by translating the temporal information of single photons to a wavelength distribution of frequency conversion followed by de-multiplexed detection. It uses a multi-color, pulse-delayed pump laser to drive multiplexed frequency conversion, transducing photons to various frequency channels according to their arrival time. By photon detection in each channel, the measurement resolution and accuracy can reach picosecond level, much lower than the detectors' naive resolution and significantly beating the shot-noise limited direct detection. Distinct to any method relying on repeated, multiple sampling, our approach supports event-ready operations, capable of detecting randomly arriving single photons with no dead window. It is thus particularly suitable for practical applications of ranging, sensing, and communications in the dynamic, photon-starving environment. 
\end{abstract*}

\section{Introduction}

Precise measurement of single-photon arrival times is a foundational capability for emerging quantum technologies in communication, computing, ranging, and sensing \cite{hadfield2009single,Eisaman2011,becker2005TCSPC,rehain2020,rapp2021}. For many applications, timing resolution directly impacts overall system performance, determining information capacity, measurement precision, processing speed, and more \cite{becker2005TCSPC,dalla2020time,scholes2023fundamental,hirvonen2020fast,perenzoni2016compact}. The past decade has witnessed significant technological advances in Si and In-GaAs avalanche photodiodes, transition edge sensors, and superconducting nanowire single-photon detectors  \cite{natarajan2012,zhang2015InGaAs,korzh2020}, as well as in their accompanying time-correlated single-photon counting (TCSPC) and time-tagging electronics \cite{becker2005TCSPC,stevens2006,farina2021}. However, fundamental trade-offs remain among detection efficiency, timing jitter, and dark counts \cite{natarajan2012,zhang2015InGaAs,bruschini2019single}. In particular, single-photon detection with high efficiency, low noise, short dead time and picosecond or better resolution remains a significant challenge, mainly because it requires ultrafast electronics \cite{esmaeil2021superconducting,korzh2020,farina2021}.

A viable solution is to circumvent the need for ultrafast electronics by replacing them with fast optics, thereby combining the complementary advantages of electronic and optical processing \cite{kolner1989temporal,huang2022wide,joshi2022picosecond}. To this end, here we demonstrate ultrafast photon detection by substituting the otherwise expensive and challenging terahertz electronic time tagging with picosecond optical time gating. The gating is realized through frequency conversion driven by picosecond pulses, such that the signal can be converted only when it arrives within the picosecond window of a pump pulse \cite{kuzucu2008time,eckstein2011}. This optical gating technique is not new and has previously been demonstrated in ultrafast upconversion-based single-photon detection and imaging systems \cite{kuzucu2008time,huang2022wide,fang2023mid}. However, these approaches require an optical delay line to sweep the pump pulses to capture the signal \cite{huang2022wide,fang2023mid,yabuno2022ultrafast}. Hence, they are inherently slow and do not support event-ready detection, where a transient signal may arrive at an arbitrary and unknown time. 

In contrast, the new technique we demonstrate here uses a set of frequency-interleaved and relatively delayed pump pulses to drive multiplexed frequency conversion that is simultaneously phase matched \cite{tang2024mQFC,serino2023,velev2014selective}. The pulse set consists of relatively delayed pulses of different wavelengths arranged adjacently in time. Whenever a signal photon arrives, one or more pump pulses are ready to ``catch'' it and convert it to one or more new wavelengths for subsequent detection \cite{eckstein2011,brecht2011,reddy2017,allgaier2017}. Its precise arrival-time information is then translated to the converted wavelength(s), so that picosecond measurement resolution is achievable using picosecond pump pulses, which can be routinely generated by mode locking or line-by-line waveform shaping \cite{cundiff2010OAWG,huang2008,jiang2005,weiner2011,willits2012}. This realizes a single-photon measurement technique with picosecond timing resolution without using terahertz electronics. In addition, it eliminates the need for optical delay scanning, allowing event-ready transient measurement \cite{huang2022wide,fang2023mid}. 

Multiplexed conversion can be achieved by passing the signal subsequently through multiple converters, each supporting a single conversion frequency, or by using a single converter that supports multiple phase-matching conditions simultaneously \cite{tang2024mQFC,serino2023,velev2014selective}. The single-device approach is more efficient because it minimizes insertion loss. An example is a periodic poled lithium niobate (PPLN) waveguide with interleaved multiple poling periods \cite{tang2024mQFC,velev2014selective}. At the output, the converted photons are frequency-demultiplexed using, e.g., a grating filter, followed by individual detectors. The distribution of photon counts in each converted frequency channel thereby contains the information of signal arrival time, because the conversion efficiency is determined by the overlap of the signal with each pump pulse train \cite{eckstein2011,brecht2011,reddy2017}. By using picosecond or shorter pulses, the timing resolution can reach the picosecond level or higher. In this design, there is no need for ultrahigh-speed detectors or time-tagging electronics. Instead, they only need to be faster than the repetition rate of each train to eliminate ambiguity \cite{becker2005TCSPC,stevens2006,farina2021,albota2004,langrock2005,pelc2011,pelc2012}. 

Our technique is based on recent progress in quantum frequency conversion and its broad applications in quantum computing \cite{eckstein2011,brecht2011,brecht2015photon,reddy2017,reddy2018,allgaier2017}, quantum networking \cite{tanzilli2005, mcguinness2010,rakher2010,zaske2012}, and sensing \cite{shahverdi2017,garikapati2023programmable,rehain2020}. Although multiplexed frequency conversion has been demonstrated previously \cite{tang2024mQFC}, the unique single-shot capability introduced here is new and particularly suitable for field-deployable applications featuring highly dynamic, random, and non-repeating photon registration events, where multi-shot accumulation is infeasible. Key applications include remote sensing, such as satellite or airborne single-photon LiDAR for Earth observation, target tracking, and photon-efficient ranging \cite{hadfield2023single,degnan2024evolution,hong2024airborne}, and space-based quantum optical communications, where individual photons arrive unpredictably due to orbital dynamics, atmospheric variability, link motion, or protocol constraints, allowing only single-event timing estimation \cite{liao2017satellite,yin2017satellite,sidhu2021advances}.

\section{Concept}
Our idea is inspired by pulse stretching, where the spectral spread is translated into temporal spread as the pulses propagate in a dispersive medium \cite{agrawal1989ultrashort,diels2006ultrashort,godin2022recent,mahjoubfar2017time,joshi2022picosecond}. Applied inversely, if photons arriving at different times can be mapped onto different spectra, they can easily be distinguished by wavelength de-multiplexing followed by detection. However, unlike pulse stretching, this cannot be achieved using linear optics alone, and nonlinear optics must be employed \cite{kolner1989temporal,azana2003time,driouche2023tunable}.

Our goal is to develop a robust photon-detection technique with high timing resolution for a variety of practical applications. As such, we require it to satisfy
\begin{itemize}
    \item Low background noise, where the injected noise is comparable to or lower than the detector dark counts;
    \item Single-shot operation, where the system is capable of precisely determining a random photon arrival time without relying on repetitive or adaptive measurements.  
\end{itemize}
These requirements are essential for applications in photon-starved and transient environments. To this end, we designed our system as sketched in Fig.~1. It uses a multiplexed quantum frequency converter (m-QFC) that supports frequency conversion between a single signal wavelength and multiple pump wavelengths. This type of m-QFC can be realized in a periodically poled $\chi^{(2)}$ waveguide or crystal with multiple poling periods \cite{tang2024mQFC}. The m-QFC is pumped by multiple pulse trains, each of a different color, to convert the signal into different wavelengths through the generation of sum- or difference-frequency. All pulse trains share the same period, but are relatively delayed with respect to one another. Such pulse trains can be generated with frequency-comb lines, where each train is formed by a subset of comb lines. The period is determined by the comb-line spacing, whereas the pulse width is determined by the spectral width of the subset. The m-QFC output then passes through one or more frequency-demultiplexing devices, where the different converted wavelengths are separated and detected individually by single-photon detectors. In this way, the detectors provide the coarse photon arrival time, limited by their timing jitter, while the identity of the detector that clicks provides the fine timing information, determined by the pump pulse width.

In this study, we explore an implementation by combing programmable optical waveform synthesis, multiplexed nonlinear frequency conversion, and conventional single-photon detection with time-tagging. The system consists of a frequency-comb-based optical arbitrary waveform generator (OAWG) and a multi-period PPLN waveguide. OAWG synthesizes structured pump and signal pulses through line-by-line amplitude and phase control \cite{cundiff2010OAWG,huang2008,jiang2005,weiner2011,willits2012}, while the PPLN waveguide generates multiplexed sum-frequency signals in several spectrally separated visible channels \cite{tang2024mQFC,shahverdi2017}. Line-by-line pulse shaping enables deterministic control of individual comb lines, supporting user-defined ultrafast waveforms and multi-wavelength pump synthesis \cite{cundiff2010OAWG,huang2008,jiang2005,weiner2011,willits2012}. Programmable comb shaping has been demonstrated across large numbers of lines and in both fiber-laser and microresonator-comb platforms, providing a mature toolkit for coherent waveform engineering \cite{jiang2007NatPhoton,ferdous2011NatPhoton,okawachi2011octave}. This capability is essential here because the delayed multicolor pump structure described in the concept section is generated directly through programmable comb-line selection and phase control.

We realize the m-QFC in a multi-period PPLN waveguide engineered to support several independently phase-matched sum-frequency generation (SFG) channels within a single integrated device. Recent multiplexed-QFC demonstrations have shown that such channels can be engineered and addressed using multiple poling periods together with appropriate pump selection \cite{tang2024mQFC,serino2023}. In our architecture, each channel $i$ yields a detected photon count $N_i(\tau)$ that depends on the relative delay $\tau$ between the signal and the pump pulses. This delay dependence arises from the channel-specific temporal overlap between the pump and signal envelope, and the corresponding phase-matching response. Conceptually, this mechanism parallels the selective QFC and quantum pulse-gate strategies, in which pump engineering determines which temporal modes are efficiently converted \cite{eckstein2011,brecht2011,brecht2015photon,reddy2017,reddy2018,allgaier2017}. Here, however, the same principle is used to encode the photon arrival time into a multichannel spectral distribution that can be read out with standard detectors.

In a practical example, the m-QFC can consist of a PPLN waveguide with 10 quasi-phase-matching peaks \cite{velev2014selective}. Ten pulse trains are carved out of a laser beam of comb lines with $25$-GHz spacing; see Fig. 1(a). Each pulse is about 4 ps long, so within each period, there are 10 pulses of different colors that span the full period of each pulse train. There are 10 single-photon detectors, each assigned to capture the converted signal at a different wavelength. The timing jitter of commercially available Si-APD detectors is typically about 40 ps FWHM. Therefore, when a photon is detected, the detector gives the coarse photon arrival time to determine which period it falls. The which-detector information, on the other hand, gives a fine resolution of 4 ps.

\begin{figure}[H]
\centering
\begin{adjustbox}{width=0.99\textwidth, height=0.35\textheight, keepaspectratio}
\input{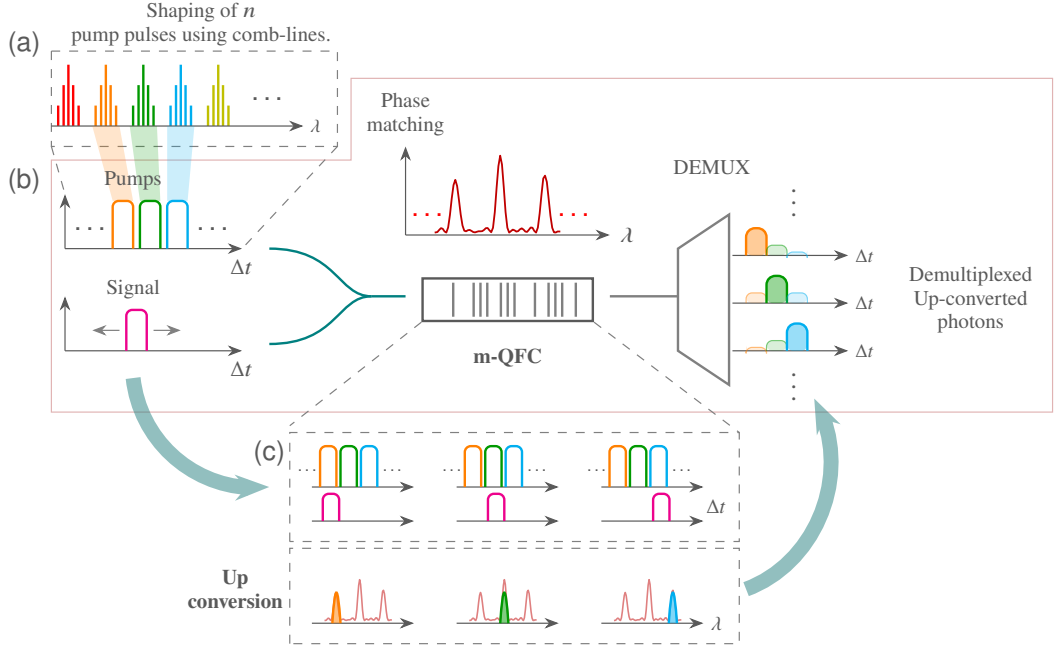}
\end{adjustbox}
\caption{\textit{(a)} Representative pulse trains from the comb-lines. Conceptual diagram illustrating how a relative delay alters the temporal overlap within the m-QFC stage \textit{(b,c)}, resulting in the redistribution of up-converted photons among output channels.}

\end{figure}

There are multiple ways to infer the photon arrival time from the photon counting results in each channel. Here, we use a ratio-manifold estimator that compares the measured spectral count distribution with a calibrated delay-dependent reference. For each calibrated delay $\tau$, the normalized ratio in the $i$-th conversion channel is defined as
\begin{equation}
R_i(\tau)=
\frac{N_i(\tau)}
{\sum_{j=1}^{M}N_j(\tau)},
\qquad i=1,\dots,M,
\end{equation}
where $M$ is the number of conversion channels, \(N_i(\tau)\) is the photon count in channel \(i\) at delay \(\tau\), and \(\sum_{j}N_j(\tau)\) is the total detected count over all channels. This normalization suppresses common-mode power fluctuations and channel-independent throughput variations, while preserving the relative inter-channel pattern that encodes the timing information. For an unknown photon-arrival time, the measured counts \(n_i\) are first normalized to form the ratio vector 
\begin{equation}
r_i=\frac{n_i}{\sum_{j=1}^{M} n_j},
\qquad i=1,\dots,M.
\end{equation}
The measured and calibrated ratio vectors are then written as
\begin{equation}
\mathbf{r}_{\mathrm{meas}}=
[r_1,\dots,r_M]^{\mathsf{T}},
\qquad
\mathbf{R}(\tau)=
[R_1(\tau),\dots,R_M(\tau)]^{\mathsf{T}}.
\end{equation}

The arrival time is estimated by comparing \(\mathbf{r}_{\mathrm{meas}}\) with the calibrated manifold over the delay set \(\mathcal{T}\). Specifically, we define the distance function
\begin{equation}
D(\tau)=\left\|\mathbf{R}(\tau)-\mathbf{r}_{\mathrm{meas}}\right\|_2^2,
\end{equation}
and determine the estimated photon arrival time as
\begin{equation}
\hat{\tau}=\arg\min_{\tau\in\mathcal{T}} D(\tau).
\end{equation}

In the high-count regime, the ratio mapping approaches a deterministic manifold, whereas at low photon number, the estimation uncertainty is governed by shot-noise fluctuations in the underlying counts \cite{MandelWolf}. The detection stage employs standard single-photon detectors, for example silicon single-photon avalanche photo-diodes (Si-APDs) following up-conversion, together with commercial time-tagging electronics, thereby leveraging established multichannel photon-counting platforms \cite{albota2004,langrock2005,pelc2011,pelc2012,becker2005TCSPC,farina2021}. Because the estimator operates entirely in calibrated ratio space, sub-picosecond reporting can be achieved by evaluating the manifold on a finer delay grid without modifying the physical hardware. The resulting architecture therefore unifies programmable comb-based waveform synthesis, multiplexed nonlinear conversion, and calibration-based statistical inversion into a scalable platform for photon arrival-time estimation using standard detectors and readout electronics \cite{cundiff2010OAWG,tang2024mQFC,becker2005TCSPC}.

It is worth noting that this estimator is perhaps the simplest method to find the arrival time, for illustration purpose. More sophisticated methods, including those of machine learning, could give even higher resolution and/or accuracy. Also, it does not utilize any time-tag information from the detectors. Such information, while coarser than what we can derive from the pump pulse width with low photon counts, can be finer with high photon counts, by virtue of shot noise. A better data processing method can improve the system performance and will be a subject of future studies. 

\section{Experimental setup}

The experimental configuration, shown in Fig.~2, consists of three stages: (i) OAWG, (ii) nonlinear frequency conversion, and (iii) wavelength-resolved single-photon detection.

In the first stage, a tunable laser source (TLS) provides a continuous-wave input to an optical frequency comb generator (OFCG, WTEC-01-25) through a fiber polarization controller (FPC). The OFCG is temperature stabilized at $50^{\circ}\mathrm{C}$ and driven by a $25~\mathrm{GHz}$ RF signal, producing a broadband comb with $25~\mathrm{GHz}$ line spacing. The comb output ($-12~\mathrm{dBm}$) is amplified by an erbium-doped fiber amplifier (EDFA, $23~\mathrm{dB}$ gain) and split using a 90:10 splitter, with the $10\%$ port monitored by an optical spectrum analyzer (OSA) to verify spectral stability. The remaining power is directed to a $10~\mathrm{GHz}$ resolution wave-shaper (Finisar 16000A), which performs line-by-line amplitude and phase control to synthesize the pump and signal wavelengths.

In the second stage, following pulse synthesis, EDFAs compensate the wave-shaper insertion loss ($\sim 4.5~\mathrm{dB}$). Fiber polarization controllers and a fiber polarizer (pump arm) prepare the polarization state for nonlinear interaction. An optical delay line in the signal arm controls the relative delay $\tau$ between the pump and the signal pulses. The pump arm includes $10~\mathrm{m}$ single-mode fiber to introduce controlled chromatic dispersion and generate relative timing offsets between comb lines. The pulses are then coupled into a multi-PPLN waveguide that supports the multiplexed SFG.

In the third stage, the generated visible outputs are separated using a free-space spectrograph based on a diffraction grating. Steering mirrors direct the dispersed beams into three spatial paths, each coupled to a fiber collimator and filtered using cascaded $1~\mathrm{nm}$ band-pass filters for spectral isolation. Photon arrival events are detected using three Si-APDs and recorded using a synchronized time-tagging unit.

Channel-dependent throughput variations arising from coupling and filtering are compensated in post-processing through the use of normalized photon-count ratios rather than absolute count rates.

\begin{figure}[H]
\centering
        \begin{adjustbox}{width=0.99\textwidth, height=0.4\textheight, keepaspectratio}
        \input{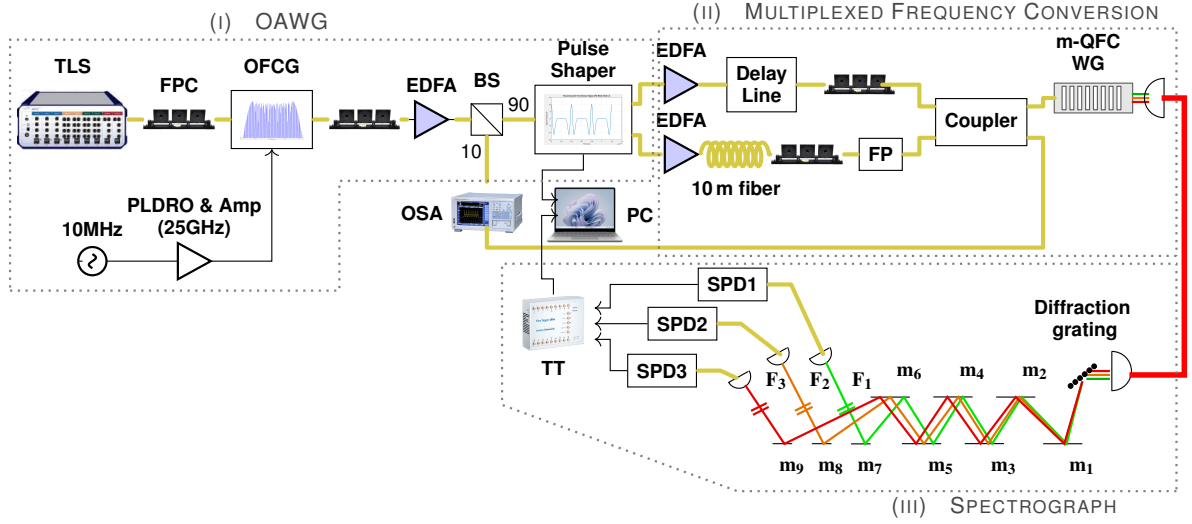}
        \end{adjustbox}
        \caption{Experimental platform for time-to-frequency multiplexed photon arrival-time estimation. Stage (i) comb-based OAWG synthesizes the pump and signal pulses; stage (ii) a multi-period PPLN waveguide performs multiplexed frequency conversion; stage (iii) a grating spectrograph de-multiplexes three SFG channels for single-photon detection and time tagging. EDFA: erbium-doped fiber amplifier; OFCG: optical frequency comb generator; OSA: optical spectrum analyzer; BS: beam splitter; FP: fiber polarizer; $m_i$: mirrors; $F_i$: filters; TT: time tagger.}
\end{figure}

\section{Results and Discussion}

We first verify the spectral operating window of the multiplexed nonlinear converter before evaluating its timing performance. Waveguide characterization at $39.4 \pm 0.1^{\circ}\mathrm{C}$ reveals three distinct second harmonic generation (SHG) phase-matching peaks at $1552.7~\mathrm{nm}$, $1555.2~\mathrm{nm}$, and $1557.7~\mathrm{nm}$, each corresponding to the three poling periods. Figure~3 summarizes the phase-matching map and identifies the corresponding SHG peaks (stars) together with the three operating SFGs (squares). For timing measurements, a fixed telecom-band signal at $1560.666~\mathrm{nm}$ interacts with pump wavelengths of ($1544.775~\mathrm{nm}$, $1549.575~\mathrm{nm}$, and $1554.385~\mathrm{nm}$), producing spectrally separated visible outputs at $776.34~\mathrm{nm}$, $777.55~\mathrm{nm}$, and $778.76~\mathrm{nm}$, respectively. The signal and pump pulses have intensity full width at half maximum (FWHM) pulse widths of approximately $3.8~\mathrm{ps}$ and $12.5~\mathrm{ps}$, respectively. Because the conversion efficiency in each channel depends on the pump--signal temporal overlap, the measured photon counts across the different channels vary with delay. These delay-dependent count patterns form a multichannel signature that encodes the arrival time $\tau$.

\begin{figure}[H]
\centering
\includegraphics[width=0.6\textwidth,height=0.25\textheight]{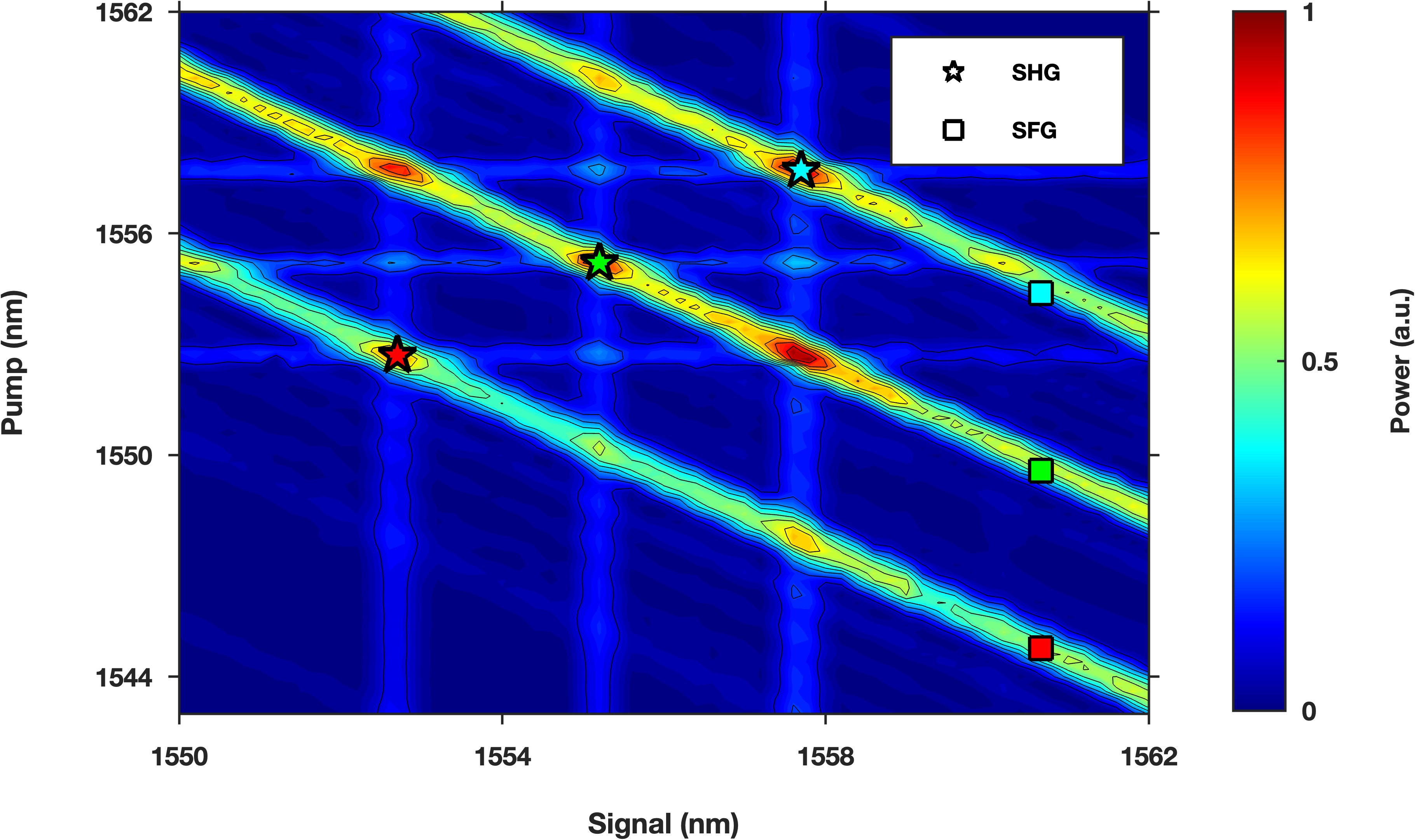}
\caption{Phase matching map of the multi period m-QFC PPLN waveguide showing SHG (stars) and SFG (squares) peaks for three engineered poling periods, enabling simultaneous multichannel frequency conversion.}
\end{figure}

To characterize the temporal response of the m-QFC, we scanned the relative delay $\tau$ between the signal and pump pulses and recorded the photon counts in each wavelength-converted channel. Figure~4 shows the delay-dependent counts $N_{1,2,3}(\tau)$ for integration times of $100~\mathrm{ms}$ and $5~\mathrm{ms}$. Each channel exhibits a distinct temporal response peak, reflecting the channel-dependent pump--signal overlap inside the multi-period PPLN waveguide. As the delay is varied, the signal photon is therefore converted with different efficiencies into the three output wavelengths, producing a multichannel temporal signature that forms the basis of the arrival-time estimator.

\begin{figure}[H]
\centering
\includegraphics[width=0.9\textwidth,height=0.25\textheight]{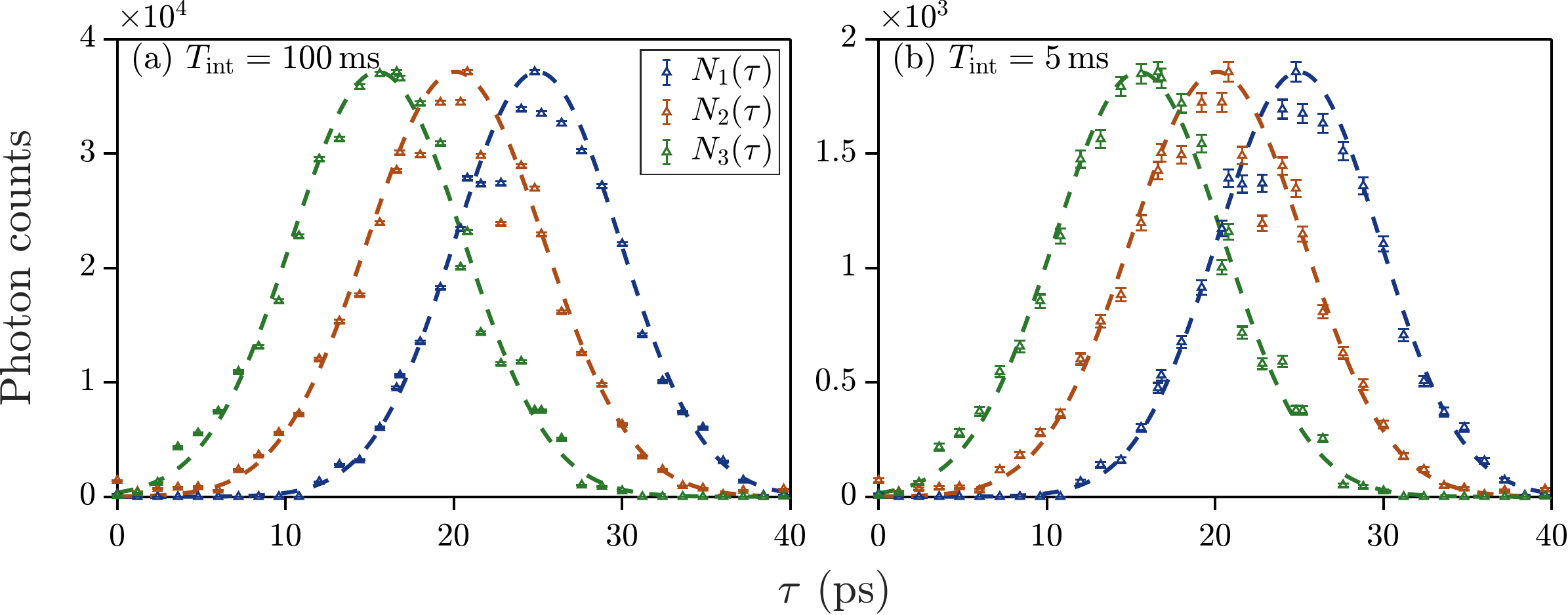}
\caption{Delay-dependent photon counts in the three wavelength-converted channels for (a) $T_{\mathrm{int}}=100~\mathrm{ms}$ and (b) $T_{\mathrm{int}}=5~\mathrm{ms}$. Symbols denote measured counts with Poisson uncertainty; dashed curves are Gaussian fits. The separated response peaks demonstrate stable time-to-frequency mapping across the three conversion channels.}
\end{figure}

At $T_{\mathrm{int}}=100~\mathrm{ms}$, shown in Fig.~4(a), the photon-count fluctuations are suppressed by averaging, each channel shows a smooth and well-separated temporal profile. The dashed Gaussian fits are used only as visual guides to identify the peak positions and widths of the channel responses; the subsequent arrival-time retrieval is performed using the calibrated photon-count ratios rather than the fitted curves. As shown in Fig.~4(b), when the integration time is reduced to $T_{\mathrm{int}}=5~\mathrm{ms}$, the same peak structure is preserved, although the shot-noise fluctuations become more visible. The error bars represent the expected Poisson counting uncertainty ($\sigma=\sqrt{N}$), confirming that the observed fluctuations are consistent with photon-counting statistics.
 
The persistence of three separated response peaks at (5) ms demonstrates that the nonlinear time-to-frequency mapping remains stable under reduced photon number. This result is important because it shows that the optical conversion stage provides sufficient channel contrast for ratio-based arrival-time inference in the millisecond integration regime. For clarity, the delay axis is shifted as $\tau\rightarrow\tau-20~\mathrm{ps}$, mapping the experimental scan range of 20--60~$\mathrm{ps}$ to the displayed window of 0--40~$\mathrm{ps}$ within a period of comb lines.

\begin{figure}[H]
\centering
\includegraphics[width=0.9\textwidth,height=0.4\textheight]{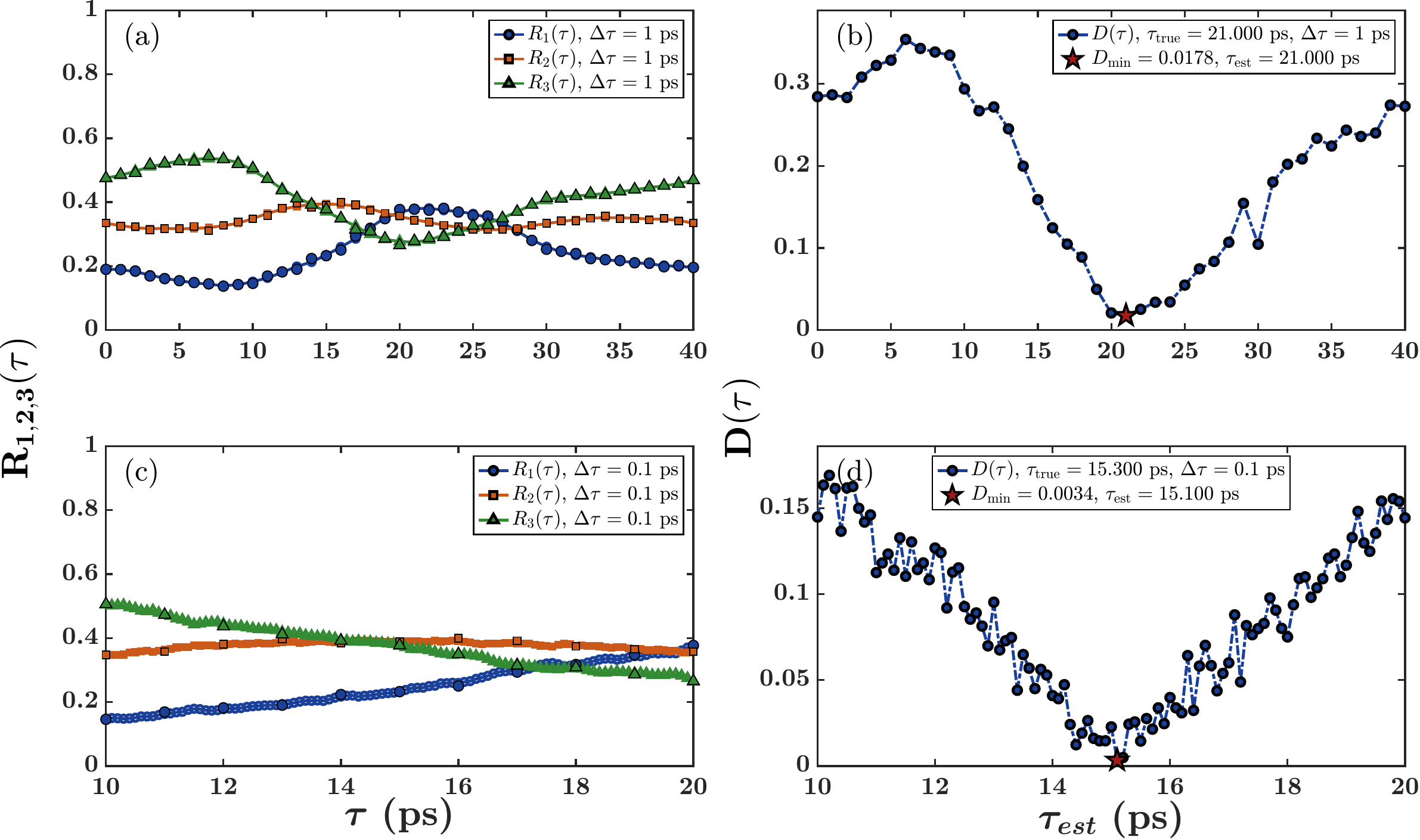}
\caption{Ratio manifold calibration and distance-based arrival-time estimation at $T_{\mathrm{int}}=100~\mathrm{ms}$. 
(a) Calibrated channel ratios $(R_{1,2,3}(\tau))$ over the full delay window using a $\Delta\tau=1~\mathrm{ps}$ grid. 
(b) Distance function $(D(\tau))$ for a representative measurement at $\tau_{\mathrm{true}}$; the red star marks the minimum, which gives the estimated arrival time $\tau_{\mathrm{est}}$. 
(c) Zoomed-in calibrated ratios using a finer ($\Delta\tau=0.1~\mathrm{ps}$) grid. 
(d) Corresponding fine-grid distance function, where the starred minimum gives $\tau_{\mathrm{est}}$.}
\end{figure}

To infer the photon arrival time, the channel resolved counts are converted into normalized photon-count ratios and compared with the calibrated ratio manifold. Figure~5 shows this procedure at $T_{\mathrm{int}}=100~\mathrm{ms}$ for two calibration-grid spacings. As shown in Fig.~5(a), the ratios ($R_i(\tau)$ ($i=1,2,3$)) are sampled over the full delay window using a coarse grid of $\Delta\tau=1~\mathrm{ps}$. The three ratio curves vary smoothly with delay, providing a unique multichannel signature for arrival-time estimation. For a measurement at an unknown delay, the measured ratio vector is compared with each point on the calibrated ratio manifold by evaluating the distance function ($D(\tau)$), as shown in Fig.~5(b). The minimum of $D(\tau)$, marked by the red star, identifies the best-matched calibration point and gives the estimated arrival time ($\tau_{\mathrm{est}}$). In this case, the minimum occurs at $\tau_{\mathrm{est}}=21.000~\mathrm{ps}$, correctly recovering the applied delay on the $1~\mathrm{ps}$ grid. 
The same method is applied to a finer calibration grid of $\Delta\tau=0.1~\mathrm{ps}$ over a zoomed delay interval, as shown in Fig.~5(c,d). The finer grid gives the estimator more possible delay values and can therefore reduce the error associated with a coarse ($1~\mathrm{ps}$) sampling step. However, neighboring ($0.1~\mathrm{ps}$) calibration points are closer together in ratio space, making them harder to distinguish when photon-counting noise is present. Therefore, the finer grid is more sensitive to photon-counting noise under low-count conditions. Nevertheless, the distance curve shown in Fig.~5(d), retains a clear minimum, showing that sub-picosecond retrieval is possible when the photon counts are sufficient to resolve the closely spaced calibration points.

\begin{figure}[H]
\centering
\includegraphics[width=0.9\textwidth,height=0.4\textheight]{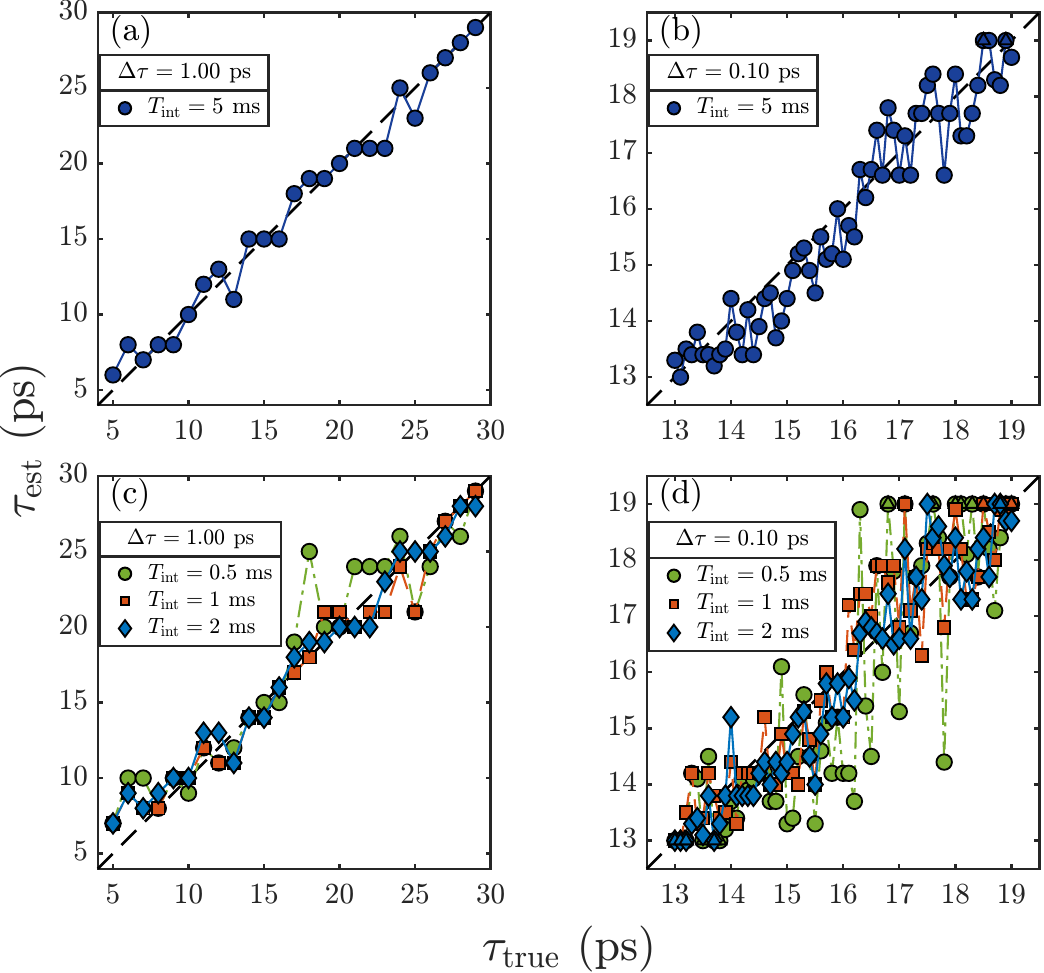}
\caption{True versus estimated arrival time obtained using the photon ratio minimum distance estimator. Panels (a,c) show coarse grid retrieval ($\Delta\tau=1~\mathrm{ps}$), while panels (b,d) show dense sub-ps retrieval ($\Delta\tau=0.1~\mathrm{ps}$) over a zoomed interval. 
Triangles denote estimates clipped at the axis limits in the zoomed panels. 
Integration times $T_{\mathrm{int}}$ are indicated in each panel. 
The dashed line marks ideal agreement, $\tau_{\mathrm{est}}=\tau_{\mathrm{true}}$.}
\end{figure}

Figure~6 evaluates the retrieval accuracy of the ratio-based estimator by comparing the estimated photon arrival time ($\tau_{\mathrm{est}}$) with the applied delay time ($\tau_{\mathrm{true}}$). The dashed diagonal represents ideal recovery, and deviations from this line directly indicate the estimator error. The retrieval error is quantified by the root-mean-square error (RMSE),
$
\mathrm{RMSE}
=
\sqrt{
\frac{1}{N}
\sum_{k=1}^{N}
\left(
\tau_{\mathrm{est},k}
-
\tau_{\mathrm{true},k}
\right)^2
},
$
where $N$ is the number of evaluated delay points. This metric directly measures how accurately the calibrated wavelength-channel ratio vector recovers the photon arrival time. Figure~6(a,c) shows the retrieval performance using a coarse calibration grid with $\Delta\tau=1~\mathrm{ps}$. At $T_{\mathrm{int}}=5~\mathrm{ms}$, as shown in Fig.~6(a), the estimated photon arrival times closely follow the ideal diagonal, giving an RMSE of $1.02~\mathrm{ps}$ over the evaluated timing window. This demonstrates that the relative photon-count distribution among the three frequency-converted channels provides a stable and nearly one-to-one timing signature, enabling picosecond-scale arrival-time estimation in the millisecond integration regime. When the integration time is reduced, as shown in Fig.~6(c), the detected photon number decreases and the normalized ratios become more sensitive to shot-noise fluctuations. As a result, the retrieved points show increased scatter around the diagonal. The RMSE values are $2.29~\mathrm{ps}$, $1.41~\mathrm{ps}$, and $1.22~\mathrm{ps}$ for $T_{\mathrm{int}}=0.5~\mathrm{ms}$, $1~\mathrm{ms}$, and $2~\mathrm{ms}$, respectively. The improvement with increasing integration time indicates that the retrieval error is strongly influenced by photon-counting noise in the ratio measurement rather than by instability of the nonlinear conversion process. 

Figure~6(b,d) shows the retrieval performance using the finer calibration grid with $\Delta\tau=0.1~\mathrm{ps}$ over a zoomed timing interval. In this case, the calibrated ratio manifold is sampled more densely than in the $\Delta\tau=1~\mathrm{ps}$ retrieval, allowing the measured channel-ratio vector to be matched to more closely spaced delay values. At $T_{\mathrm{int}}=5~\mathrm{ms}$, the photon-count ratios are stable enough to support this finer search, and the retrieved timing remains close to the ideal diagonal with an RMSE of $0.57~\mathrm{ps}$, as shown in Fig.~6(b). This sub-picosecond RMSE shows that the wavelength-channel photon count ratios carry arrival-time information below the $1~\mathrm{ps}$ coarse-grid spacing and beyond the naive timing resolution of the detector and time-to-digital converter (TDC) chain. The finer grid, however, also places a stronger requirement on photon statistics. For $\Delta\tau=0.1~\mathrm{ps}$, neighboring calibration points produce very similar ratio vectors. Under low-count conditions, shot-noise fluctuations in the detected channel counts can therefore move the measured ratio vector toward an incorrect delay point, leading to larger retrieval errors. The behavior is evident as shown in Fig.~6(d), where the shortest integration time of $0.5~\mathrm{ms}$, gives an RMSE of $3.20~\mathrm{ps}$. As the integration time increases to $1~\mathrm{ms}$ and $2~\mathrm{ms}$, more photons are collected, the ratio fluctuations are reduced, and the RMSE improves to $0.89~\mathrm{ps}$ and $0.62~\mathrm{ps}$, respectively. This trend shows that the fine-grid estimator becomes reliable when the photon-counting noise is low enough to distinguish adjacent points on the calibrated ratio manifold.

Together, these results show the balance between calibration-grid spacing and photon-counting noise. The $1~\mathrm{ps}$ grid gives robust picosecond-scale retrieval under lower-count conditions because neighboring calibration points are more clearly separated in ratio space. The $0.1~\mathrm{ps}$ grid reduces the finite-grid error and enables sub-picosecond retrieval, but only when the detected photon number is high enough to distinguish the closely spaced ratio patterns. Therefore, the measured RMSE values quantify the retrieval accuracy of the ratio-manifold estimator and identify the photon-counting conditions required for stable high-resolution operation.

\begin{figure}[t]
\centering
\setlength{\figw}{0.95\linewidth}

\begin{tikzpicture}[x={\dimexpr\figw/532\relax},y={\dimexpr\figw/532\relax}]
  \node[anchor=south west, inner sep=0pt] at (0,0)
    {\includegraphics[width=\figw]{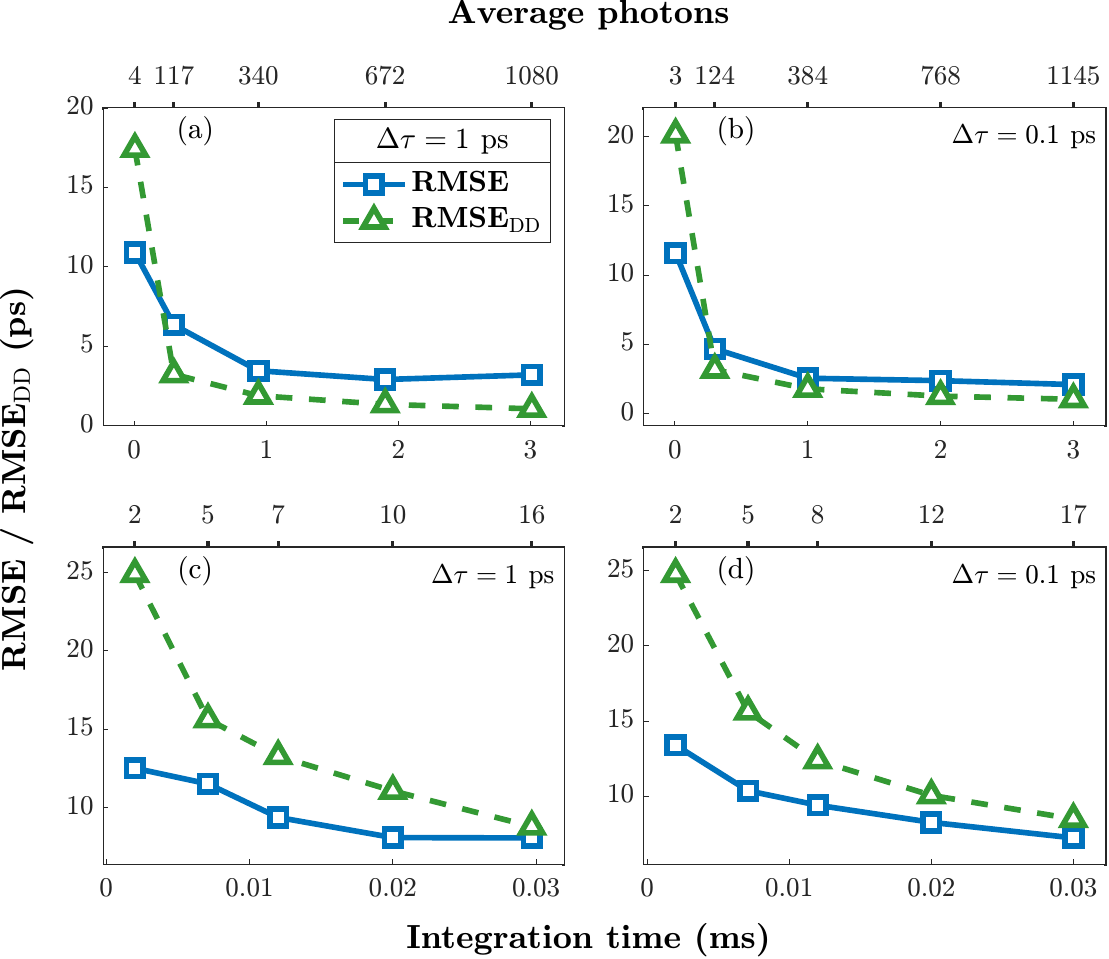}};

  \shadeband{1.22*\PanelLx}{0.6*\PanelLxr}{\TopYb}{\TopYt}{0}{\bandfrac}
  \shadeband{1.035*\PanelRx}{0.8*\PanelRxr}{\TopYb}{\TopYt}{0}{\bandfrac}

  \shadeband{\PanelLx}{\PanelLxr}{\BotYb}{\BotYt}{0}{1}
  \shadeband{\PanelRx}{\PanelRxr}{\BotYb}{\BotYt}{0}{1}

  \zoomcol{1.2*\PanelLx}{\PanelLxr}{\TopYb}{\BotYt}{50}{50}
  \zoomcol{1.03*\PanelRx}{\PanelRxr}{\TopYb}{\BotYt}{312}{308}
\end{tikzpicture}

\caption{Timing RMSE versus integration time for the proposed estimator compared
with a direct-detection reference. Panels (a,c) use a \(1~\mathrm{ps}\) calibration
grid, and panels (b,d) use a \(0.1~\mathrm{ps}\) grid; panels (c,d) zoom into the photon-starved regime. Blue squares show the measured estimator RMSE, while green
triangles show the direct-detection reference, using the measured detector--TDC
jitter and background-subtracted net photon number. The top axis gives the mean
net photons per integration window. \emph{Shaded region}: photon-starved regime
bounded by the red noise-floor curve.}
\label{fig:timing_rmse}
\end{figure}

Figure~7 benchmarks the proposed time-to-frequency arrival-time estimator against a conventional direct-detection reference. The direct-detection benchmark is calculated from the independently measured timing jitter of the detector and  TDC chain. The direct-detection timing precision is taken as

\[
\mathrm{RMSE}_{\mathrm{DD}}(T_{\mathrm{int}})
=
\frac{
\sigma_{\mathrm{det+TDC}}
}{
\sqrt{
N_{\mathrm{sig}}(T_{\mathrm{int}})
}
},
\]
where \(\sigma_{\mathrm{det+TDC}}\) is the RMS timing jitter of the detector--TDC chain and $N_{\mathrm{sig}}(T_{\mathrm{int}})$ is the net signal photon number after background subtraction. This expression represents the conventional timestamp-averaging limit, where the timing uncertainty improves as \(1/\sqrt{N_{\mathrm{sig}}}\), while the absolute scale is set by the detector and TDC jitter. The measurement procedure used to determine $\sigma_{\mathrm{det+TDC}}$ is described in Appendix~A. 

As shown in Fig.~7, the blue curves shows the experimentally measured RMSE of the proposed ratio-based estimator, obtained by comparing the estimated arrival time with the known applied delay. The green curves shows the direct-detection reference calculated from the measured detector--TDC jitter and the corresponding net photon number at each integration time. The lower horizontal axis gives the integration time, while the upper horizontal axis gives the average net photon number, allowing the retrieval accuracy to be compared directly against the photon-counting condition.

The advantage of the proposed optical encoding is most clear in the photon-starved regime, as shown in Fig.~7(c,d), where only a few to tens of photons are detected within each integration window. In this regime, conventional direct detection is strongly limited by the small number of timestamps available for averaging. By contrast, the proposed estimator uses the wavelength-channel photon-count distribution produced by the nonlinear converter, so the fine timing information is extracted from the relative counts across the converted channels. As a result, the measured RMSE of the proposed estimator remains below the direct-detection reference over the plotted low-count range. This confirms that time-to-frequency mapping can provide a timing advantage when the photon number is too small for effective timestamp averaging.

At larger photon numbers, shown in Fig.~7(a,b), both the proposed estimator and the direct-detection reference improve as more photons are collected. The remaining error of the proposed estimator is likely set by a combination of calibration-grid spacing, photon-counting noise, residual background, channel imbalance, and calibration mismatch. These technical limits can be reduced by improved calibration, higher conversion efficiency, better channel balancing, and a larger number of wavelength channels.

Importantly, the present analysis uses only the wavelength-channel photon counts. The electronic time tags recorded by the detectors are not included in the ratio-based estimator. Therefore, the results in Fig.~7 report the performance of the optical time-to-frequency encoding itself. An estimator that combines the wavelength-channel ratios with the detector time tags could further improve the timing precision by using both the optically encoded fine timing information and the detector-provided coarse timing information. Overall, these results show that the proposed method is particularly effective in photon-starved conditions, where conventional direct detection is limited by detector--TDC jitter and by the few photons available for timestamp averaging.

\section{Conclusion}

We have experimentally demonstrated a nonlinear-optical method for precise photon arrival-time measurement. In this approach, the precise arrival-time information is mapped onto multiple wavelength channels through multiplexed frequency conversion. Using a comb-driven OAWG and a multi-period PPLN waveguide, we convert the incoming signals into three spectrally separated sum-frequency channels. The relative photon counts in these channels provide a calibrated timing pattern that is used to estimate the photon arrival time with standard single-photon detectors. 

Our current results show picosecond-scale timing estimation with a few detected photon counts. Compared with a conventional direct-detection reference, the proposed method provides a clear advantage in the photon-starved regime, where timing precision is normally limited by detector and TDC jitter. This improvement comes from shifting part of the timing measurement from the electronic domain to the optical domain: the nonlinear conversion stage converts small changes in arrival time into measurable changes in the photon distribution across wavelength channels.

A key feature of this scheme is its event-ready operation. Unlike delay-scanning optical gating methods, it does not require sweeping a pulse across the arrival-time window, which could be prohibitively slow for many applications in dynamical environment. Also, it largely mitigates the dead time associated with repetitive temporal scanning and is suitable for photons that arrive randomly or only once.

These results establish multiplexed nonlinear frequency conversion as a practical route to photon arrival-time measurement beyond the native timing resolution of the detectors. With improved detection efficiency, larger channel counts, and data processing that combines both wavelength-channel ratios and detector time-tag information, this approach can be extended toward sub-picosecond timing precision. The method is well suited for photon-starved applications in ranging, sensing, synchronization, and optical communications.

\appendix
\section{Timing-jitter measurement}
\label{app:jitter}
\renewcommand{\thefigure}{A\arabic{figure}}
\setcounter{figure}{0}

The detector and time-to-digital-converter (TDC) timing jitter used for the direct-detection reference in Fig.~7 of the main text was measured independently using a trigger-referenced timing histogram. The measurement was performed with the Si-APD detector and Swabian time-tagger electronics used in the experiment. As shown in Fig.~A1, a visible mode-locked laser provided both the optical pulses incident on the Si-APD and the timing reference recorded by the time tagger. The laser trigger was recorded on one time-tagger channel, while the Si-APD output was recorded on a second channel. The time differences between the trigger and detector events were accumulated to form the detector--TDC timing-response histogram. A representative measured timing histogram is shown in Fig.~A2.

The RMS timing jitter of the detector--TDC chain was calculated directly from the measured timing histogram after background subtraction. If $t_k$ is the center of the $k$th timing bin and $n_k$ is the corresponding background-subtracted count, the mean timing reference is

\[
t_0 =
\frac{\sum_k n_k t_k}{\sum_k n_k},
\]

and the RMS timing jitter is

\[
\sigma_{\mathrm{det+TDC}} =
\sqrt{
\frac{
\sum_k n_k(t_k-t_0)^2
}{
\sum_k n_k
}
}.
\]

Using this procedure, the measured detector--TDC RMS timing jitter is

\[
\sigma_{\mathrm{det+TDC}} \approx 38.117~\mathrm{ps}.
\]

This independently measured value was used to calculate the conventional direct-detection timing reference,
\[
\mathrm{RMSE}_{\mathrm{DD}}(T_{\mathrm{int}})
=
\frac{\sigma_{\mathrm{det+TDC}}}{\sqrt{N_{\mathrm{sig}}(T_{\mathrm{int}})}}
\]

where $N_{\mathrm{sig}}(T_{\mathrm{int}})$ is the background-subtracted net signal photon number within the integration window $T_{\mathrm{int}}$. This expression represents the conventional timestamp-averaging limit, in which the timing uncertainty decreases as $1/\sqrt{N_{\mathrm{sig}}}$ while the overall scale is set by the detector--TDC timing jitter.

\begin{figure}[h]
\centering
\begin{tikzpicture}[
    >=latex,
    font=\small,
    node distance=1.25cm,
    block/.style={draw, thick, rounded corners=2pt, minimum width=1.25cm,
                  minimum height=0.4cm, align=center},
    smallblock/.style={draw, thick, rounded corners=2pt, minimum width=1.05cm,
                       minimum height=0.7cm, align=center},
    fiber/.style={
        draw=yellow!80!black,
        line width=1.5pt,
        line cap=round,
        line join=round,
        shorten <= 0.4pt,
        shorten >= 0.4pt
    },
    trig/.style={thick, dashed, -latex}
]

\node (pd) at (1.5, 1.0)
    {\includegraphics[width=1.5cm]{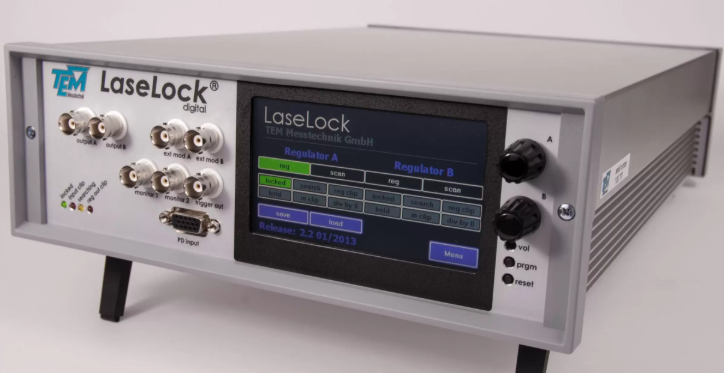}};
\node at ($(pd.north)+(0,0.3)$) {\textbf{Mod lock Laser}};

\node[block] (att) at (3.5, 0.0) {};
\node at ($(att.south)+(0,-0.3)$) {\textbf{Attenuator}};

\draw[thick, smooth, samples=50, domain=-0.35:0.35]
    plot ({\x + 2.5}, {0.42 * exp(-\x*\x / 0.005)});

\draw[thick, smooth, samples=50, domain=-0.35:0.30]
    plot ({\x + 4.55}, {0.20 * exp(-\x*\x / 0.005)});

\node (disc) at (5.55, 0.0)
    {\includegraphics[width=1.3cm]{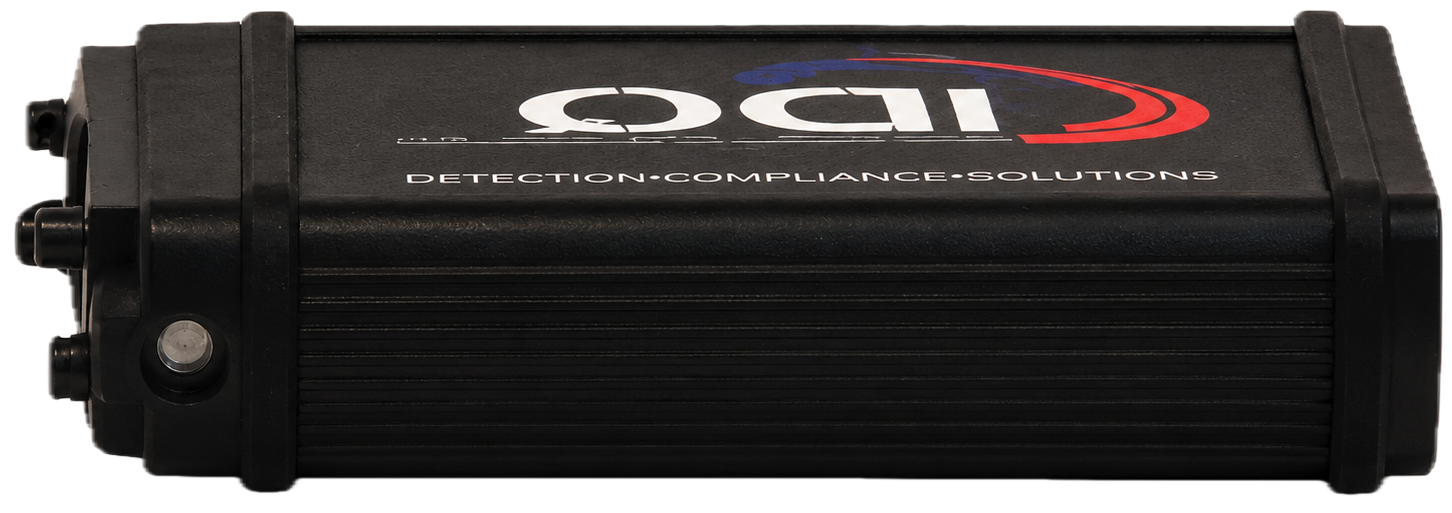}};
\node at ($(disc.south)+(0,-0.3)$) {\textbf{Si-APD}};

\node (tt) at (7.3, 0.0)
    {\includegraphics[width=1cm]{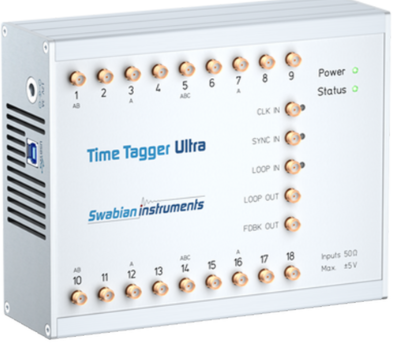}};
\node at ($(tt.south)+(0,-0.3)$) {\textbf{TT}};

\node (pc) at (9.3, 0.0)
    {\includegraphics[width=1cm]{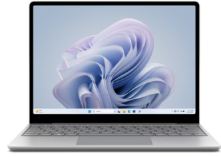}};
\node at ($(pc.south)+(0,-0.3)$) {\textbf{PC}};

\draw[fiber] (pd.south)  |- (att.west);   
\draw[fiber] (att.east)  -- ($(disc.west)+(0.1,0)$);  

\draw[->] ($(disc.east)+(-0.2,0)$) -- ($(tt.west)+(0.1,0)$);   
\draw[->] ($(tt.east)+(-0.1,0)$)   -- ($(pc.west)+(0.2,0)$);   
\draw[trig] (2.3,1.0) -| (tt.north);
\node[above] at (4.5,1) {Trigger};

\draw[densely dotted, rounded corners=4pt, gray]
    (0.25,-1.05) rectangle (10.15,1.45);

\end{tikzpicture}
\caption{Trigger-referenced timing-jitter measurement setup. A visible mode-locked laser provides the optical pulses and the timing reference. The attenuated optical pulses are detected by the Si-APD, and the laser trigger and Si-APD output are recorded on separate time-tagger channels to form the timing-response histogram.}
\label{fig:jitter_setup}
\end{figure}

\begin{figure}[H]
\centering
\includegraphics[width=0.6\textwidth,height=0.3\textheight]{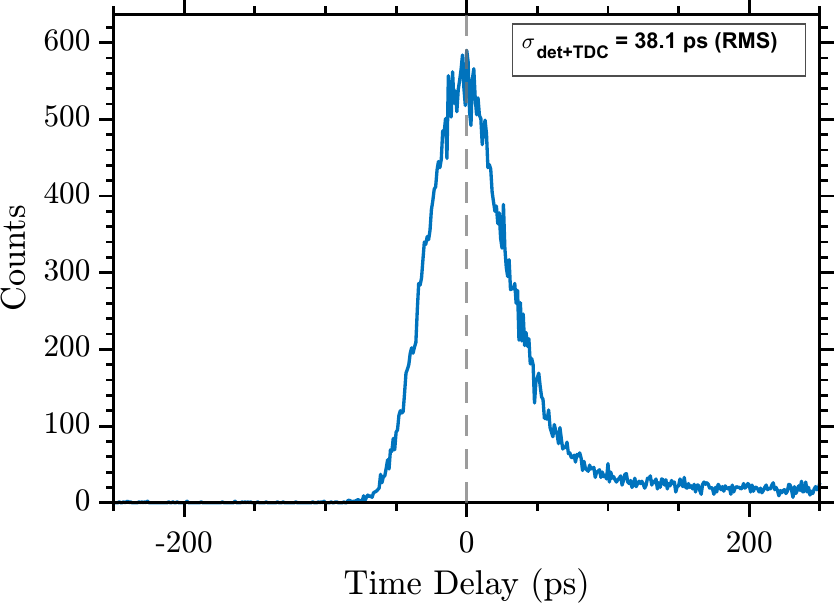}
\caption{Detector--TDC timing-response histogram measured using the trigger-referenced setup. The dashed line marks the timing reference used in the RMS calculation. The detector--TDC RMS timing jitter is $\sigma_{\mathrm{det+TDC}}=38.117~\mathrm{ps}$, which is used for the direct-detection reference in Fig.~7.}
\end{figure}

\begin{backmatter}
\bmsection{Funding}
This research was supported by ACC-New Jersey under grant number W15QKN18D0040.





\end{backmatter}

\bibliography{references}

\end{document}